\documentclass[journal]{IEEEtran}
\usepackage{amsfonts,amssymb,amsmath,algorithm,algorithmic,array}
\usepackage{bbm,bm,cite,color,dsfont,epsfig,epstopdf,graphicx}
\usepackage{setspace,subfigure,multirow,mathrsfs,multicol,xcolor}
\usepackage{amssymb}
\usepackage{amsfonts}
\usepackage{amsfonts,subfigure,multicol,color,verbatim,
graphicx,cite,epsfig,amssymb,amsmath,cases,bm,algorithm,
algorithmic,xcolor,multirow,array}
\usepackage{epstopdf}
\usepackage{booktabs}
\usepackage{indentfirst}
\usepackage{lettrine}
\begin{document}
\title{Resource Allocation in Cloud Radio Access Networks with Device-to-Device Communications}
\author{\IEEEauthorblockN{Yitao~Mo, Mugen~Peng,~\IEEEmembership{Senior Member,~IEEE,} Hongyu~Xiang, Yaohua~Sun, and Xiaodong~Ji}
\IEEEcompsocitemizethanks{
This work was supported in part by the National Natural Science Foundation of China under Grant No. 61222103, the National High Technology Research and
Development Program of China under Grant No. 2014AA01A701, and the Beijing Natural Science Foundation (Grant No. 4131003). \emph{Corresponding Author}: Mugen Peng.}
\thanks{Yitao~Mo (e-mail: moyitao\_wspn@bupt.edu.cn), Mugen~Peng (e-mail: pmg@bupt.edu.cn), Hongyu~Xiang (e-mail: xianghongyu88@163.com), Yaohua~Sun (e-mail: sunyaohua@bupt.edu.cn), and Xiaodong~Ji (e-mail: jxd@bupt.edu.cn) are with the Key Laboratory of Universal Wireless Communications (Ministry of Education), Beijing University of Posts and Telecommunications, Beijing, China.}}

\maketitle

\begin{abstract}
To alleviate the burdens on the fronthaul and reduce the transmit latency, the device-to-device (D2D) communication is presented in cloud radio access networks (C-RANs). Considering dynamic traffic arrivals and time-varying channel conditions, the resource allocation in C-RANs with D2D is formulated into a stochastic optimization problem, which is aimed at maximizing the overall throughput subject to network stability, interference, and fronthaul capacity constraints. Leveraging on the Lyapunov optimization technique, the stochastic optimization problem is transformed into a delay-aware optimization problem, which is a mixed-integer nonlinear programming problem and can be decomposed into three subproblems: mode selection, uplink beamforming design, and power control. An optimization solution that consists of a modified branch and bound method as well as a weighted minimum mean square error approach has been developed to obtain the close-to-optimal solution. Simulation results validate that the D2D can improve throughput, decrease latency, and alleviate the burdens of the constrained fronthaul in C-RANs. Furthermore, an average throughput-delay tradeoff can be achieved by the proposed solution.
\end{abstract}

\section{Introduction}

To deal with the skyrocketing increase in mobile data demands driven by data hungry applications worked on smart phones and tablets, the cloud radio access network (C-RAN) has been proposed as the evolution of ultra-dense heterogeneous wireless networks for the fifth-generation (5G) wireless network {\cite{I:cran0}}. In C-RANs, the fronthaul is used to connect the centralized processing baseband unit (BBU) pool and the distributed remote radio heads (RRHs) {\cite{I:cran1}}. The large-scale cooperative processing gains can be achieved in C-RANs because the BBU pool jointly precodes/decodes the user equipments' (UEs') symbols with centralized coordinated
multi-point (CoMP) transmission technique to improve the signal-to-interference-plus-noise ratio (SINR) {\cite{I:cran5}}. Although C-RANs have been proven to provide high spectral efficiency (SE) and energy efficiency (EE), the practical fronthaul is often capacity and time-delay constrained {\cite{I:cran2}}, which has been a significant performance bottleneck for C-RANs. To alleviate the heavy traffic burdens on the constrained fronthaul and decrease the transmit latency, device-to-device (D2D) communications can be introduced into C-RANs due to the fact that D2D communications allow the direct communication between a pair of D2D UEs of physical proximity without going through RRHs {\cite{I:cran3}}.
However, enabling D2D communications in C-RANs is challenging due to the presence of inter-tier and intra-tier interference. Without delicate designed resource allocation schemes for D2D communications, the expected gains offered by D2D communications may be counterbalanced by the severe mutual interference in C-RANs.

\subsection{Related Works}

Recently, extensive researches have been devoted to the issues related to resource allocation for D2D communications.
In {\cite{I:MS1}}, a simple power control method has been proposed for D2D communications, which limits interference between the cellular and the D2D links to constrain the SINR degradation of the cellular links.
In {\cite{I:MS2}}, a subchannel sharing scheme for D2D communications has been studied to ensure the mutual interference among the D2D pairs sharing the same subchannels is negligible.
In {\cite{I:MS3}}, to maximize the EE of D2D communications, an efficient iterative resource allocation and power control scheme for energy-efficient D2D communications underlaying cellular networks has been proposed.
The resource allocation schemes in {\cite{I:MS1}}--{\cite{I:MS3}} are done under the assumption that all D2D pairs operate in D2D mode. However, mode selection plays important roles in D2D systems since it can further improve system performance. Therefore, there have been some literatures taking mode selection into account when exploring a resource allocation solution for D2D communications.
In{\cite{I:bibi1}}, a biasing-based mode selection method for D2D-enabled single-tier
cellular networks along with truncated channel inversion power control has been investigated. The transmission mode of a D2D pair is determined by comparing the channel state information (CSI) of D2D links and cellular uplinks. In{\cite{I:bibi3}}, a dynamic stackelberg game framework has been studied, where the base station acts as the leader and all D2D UEs play as the followers, to jointly address the problem of mode selection and spectrum partitioning.
In{\cite{I:bibi4}}, three different resource sharing modes for D2D communication underlaying cellular networks have been investigated. The
optimization problem aims to maximize the sum-rate of the cellular network subject to inter-tier interference constraint.
The authors in{\cite{I:bibi2}} have proposed a joint mode selection, channel assignment, and power control algorithm to maximize
overall system throughput while guaranteeing the quality of service for both cellular and D2D links. In{\cite{I:bibi9}}, employing a distance based mode selection policy, the SE performance of D2D communications underlaying CoMP-enabled downlink C-RANs has been evaluated by using stochastic geometry.
However, the aforementioned works typically assume that all D2D pairs
are delay-insensitive without evaluating the delay performance under dynamic traffic arrivals.

In fact, a majority of proximity-based services are real-time and delay-sensitive. Moreover, schemes optimized for physical layer performance metrics, considering only the CSI, are not sufficient to ensure queue stability or packet delay requirement under the dynamic data arrivals process{\cite{I:delay1}}.
This is because the CSI only represents the transmission capabilities rather than transmission requirements. For instance, if the resource allocation scheme does not make use of the queue state information (QSI), it will hardly allocate radio resources to a D2D pair with bad channel quality but long data queue length, resulting in serious delay performance deterioration.
Towards this end, the resource allocation schemes for D2D communications should be adaptive to both the CSI and the QSI because the CSI reveals the instantaneous transmission opportunities at the physical layer and the QSI suggests the urgency of the packet flows at the media access control layer {\cite{I:delay2}}.
A dynamic power control scheme for delay-aware D2D communications under stochastic traffic arrivals has been investigated in {\cite{I:bibi5}}. Applying the queueing models, the performance of a dynamic mode selection strategy for a slot-by-slot basis D2D network has been addressed in {\cite{I:bibi6}}, which takes both random packet arrivals and fast fading channel conditions into account. An optimal dynamic mode selection and resource allocation to minimize the average delay subject to a dropping probability constraint in orthogonal frequency-division multiple-access cellular networks with D2D has been developed in {\cite{I:bibi7}}. A delay-aware algorithm to solve the problem of joint dynamic mode selection, spectrum management, power control, and interference mitigation in D2D communications underlaying LTE-A networks with both instantaneous and non-instantaneous implementations has been explored in {\cite{I:bibi8}}.
Although the delay-aware solutions in{\cite{I:bibi5}}--{\cite{I:bibi8}} can achieve significant performance improvement in conventional cellular networks with D2D, these resource allocation schemes can not be adopted to C-RANs with D2D since both the network-wide beamforming design and the impact of fronthaul capacity limitation must be explicitly taken into consideration for practical C-RANs.

\subsection{Main Contributions}

In this paper, taking queueing delay, dynamic traffic arrivals, and time-varying channel conditions into account, the resource allocation in C-RANs with D2D is formulated into a stochastic optimization problem with constraints on the network stability, interference, and
fronthaul capacity. To the best of the authors' knowledge, this is the first attempt to solve the delay-aware resource allocation optimization problem for C-RANs with D2D. To deal with this non-convex delay-aware optimization problem, the Lyapunov optimization framework is utilized to stabilize the queues of networks while maximizing the overall average throughput. The major contributions of this paper are threefold.
\begin{itemize}
\item[\emph{$\bullet$}]
The stochastic optimization problem of resource allocation is investigated for C-RANs with D2D. Different from the static optimization problems studied for underlay D2D communications in cellular networks in previous literatures, a variety of characteristics of C-RANs, such as uplink CoMP transmission technique and fronthaul capacity limitation, are considered in the proposed model. In addition, average throughput and delay are jointly incorporated into the stochastic optimization problem, which involves cross-layer optimization according to both the CSI and QSI.
\item[\emph{$\bullet$}]
As the stochastic optimization problem is a combination of instantaneous variables and time-averaged variables, the general framework of Lyapunov optimization is utilized to transform the stochastic optimization problem into the minimization of the drift-plus-penalty expression. To tackle the NP-hardness of this minimization problem with Boolean variables, it is decomposed into three subproblems: mode selection, uplink beamforming design, and power control. To decrease the computational complexity of conventional branch and bound method for mode selection, a modified version of branch and bound method is proposed. Further, a joint mode selection and resource allocation algorithm (JMSRA) based on the modified branch and bound method and weighted minimum mean square error (WMMSE) approach is proposed to solve these three subproblems iteratively without requiring any priori knowledge.
\item[\emph{$\bullet$}]
The simulation results validate that the proposed JMSRA algorithm can converge quickly and C-RANs with D2D can provide significant performance gains compared with C-RANs in terms of overall average throughput, average delay, and fronthaul consumption. In addition, the proposed JMSRA algorithm can achieve a flexible tradeoff between the average throughput and delay by adjusting the control parameter, making it simple to control the average throughput-delay performance for different kinds of applications.

\end{itemize}

\indent The rest of this paper is organized as follows. In Section \uppercase\expandafter{\romannumeral2},
the system model is introduced and the stochastic optimization
problem is formulated. In Section \uppercase\expandafter{\romannumeral3},
the general Lyapunov optimization technique
is utilized to transform the optimization
problem into a delay-aware joint mode selection and resource allocation problem, which is then iteratively solved by the proposed JMSRA algorithm. Simulation results are presented in Section \uppercase\expandafter{\romannumeral4}, followed by the conclusion in Section \uppercase\expandafter{\romannumeral5}.

\indent Throughout this paper,
lower-case bold letters denote vectors and upper-case bold letters denote matrices. $\mathbb{C}$ denotes complex domain.
The complex Gaussian distribution with mean vector $\bold{m}$ and covariance matrix $\bold{R}$ is represented by $\mathcal{CN}(\bold{m}, \bold{R})$.
$\bold{I}_M$ denotes $M$-dimensional identity matrix.
$\mathbb{E}[\cdot]$ represents expectation, while $\mathrm{Re}$$\{\cdot\}$ stands for the real part of a scalar. ${\left\|\cdot\right\|}_p$ stands for $\ell_p$-norm of a vector. The inverse, transpose, conjugate transpose are denoted as $(\cdot)^{-1}$, $(\cdot)^T$, $(\cdot)^H$, respectively. For ease of reference, the important notations used in this paper are summarized in Table \uppercase\expandafter{\romannumeral1}.

\begin{table}[!htp]
\newcommand{\tabincell}[2]{\begin{tabular}{@{}#1@{}}#2\end{tabular}}
\caption{SUMMARY OF IMPORTANT NOTATIONS USED}
\centering
\small
\begin{tabular}{|l|l|}
\hline
Symbol                 & Definition\\
\hline
$\mathcal{N}$          & The set of RRHs, defined as $\{1, 2, \cdot\cdot\cdot, N\}$ \\
\hline
$\mathcal{K}$          & The set of D2D pairs, defined as $\{1, 2, \cdot\cdot\cdot, K\}$ \\
\hline
$\textbf{v}_{n,k}$     & \tabincell{l}{The,uplink receiver beamforming vector of RRH $n$ \\for the Tx UE of D2D pair $k$}\\
\hline
$\bold{v}_{k}$         & \tabincell{l}{The \.network-wide beamforming vector, for the Tx\\ UE of D2D pair $k$}\\
\hline
$\bold{g}^{C}_{n,k}$   & \tabincell{l}{The CSI vector from RRH $n$ to the Tx UE of D2D\\ pair $k$}\\
\hline
$\bold{g}^C_k$         & \tabincell{l}{The CSI vector from \,all \,RRHs to the Tx UE of\\ D2D pair $k$} \\
\hline
$g^{D}_{i,i}$          & \tabincell{l}{The channel gain from the Tx UE of D2D pair $i$ to\\ the Rx UE of D2D pair $i$}\\
\hline
$p_k$                  & The transmit power of D2D pair $k$\\
\hline
$x_k$                  & The binary mode selection indicator of D2D pair $k$ \\
\hline
$\overline{R_k}$       & The average throughput of D2D pair $k$\\
\hline
$Q_k$                  & The data queue length of D2D pair $k$\\
\hline
$P^{I}_{Dmax}$         & \tabincell{l}{The \,tolerable interference \,threshold of D2D pairs \\operating in D2D mode} \\
\hline
$P_{max}$              & The peak transmit power of D2D pair\\
\hline
$C_n$                  & The fronthaul capacity limitation of RRH $n$ \\
\hline
\end{tabular}
\end{table}

\section{System Model and Problem Formulation}

In this section, the system model is introduced at first, then network stability and average throughput are defined, and a stochastic optimization problem is formulated at last.

\subsection{System Model}

We consider D2D communications are implemented as an underlay of uplink C-RANs with a BBU pool, $N$ RRHs, and $K$ D2D pairs, as illustrated in Fig. $1$, where each D2D pair comprises of a transmitter, named as Tx UE, and a potential receiver, named as Rx UE. Each RRH is equipped with $M$ antennas while each D2D UE is equipped with single antenna.
There are two practical transmission modes for a D2D pair, i.e., C-RAN mode and D2D mode. Specifically, the Tx UE and the Rx UE of a D2D pair operating in C-RAN mode communicate with each other through RRHs and all of RRHs can coordinately receive data symbol from the Tx UE via cooperative beamforming technique. Meanwhile, D2D pairs operating in D2D mode establish D2D links directly and reuse the same spectrum resource of C-RAN uplinks.

\begin{figure}[!htp]
\centering
\includegraphics[width=3.2in]{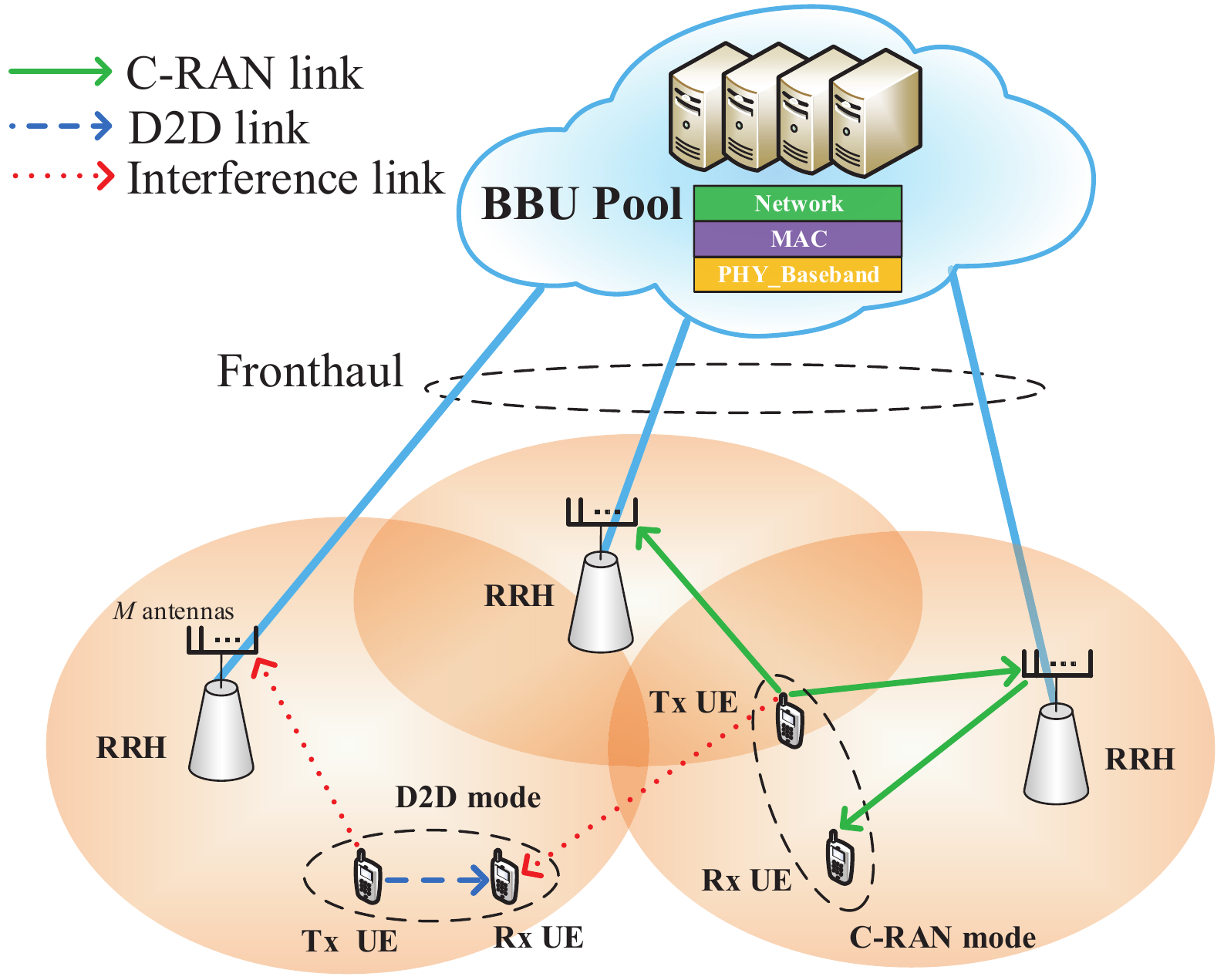}
\caption{Architecture of C-RANs with D2D.}
\label{sys}\vspace*{-1em}
\end{figure}

Assume that the network works in slotted time mode with slots
normalized to integral units, i.e., slot $t$ refers to the time interval $[t, t+1), t\in\left\{0, 1, 2, \cdots\right\}$.
Assume that the BBU pool can perfectly acquire the CSI of all C-RAN uplinks and D2D links. Furthermore, the CSI is assumed to follow quasi-static block fading, for which the channels keep constant during the duration of a slot, but identically and independently distributed (i.i.d.) over different slots. Let $\mathcal{N}=\left\{1, 2, \cdots, N\right\}$ and $\mathcal{K}=\left\{1, 2, \cdots, K\right\}$ denote the set of RRHs and the set of D2D pairs, respectively.

At slot $t$, the received signal for the Tx UE of D2D pair $k$ operating in C-RAN mode can be written as
\begin{align}
\label{UL signal}
y^{C}_k(t)= & \sum_{n=1}^{N} \bold{v}^H_{n,k}(t)\bold{g}^C_{n,k}(t)\sqrt{p_k(t)}s_k(t)\nonumber\\
&+\sum_{n=1}^{N}\sum_{\begin{subarray}{1}  l\neq k \end{subarray}}^{K}\bold{v}^H_{n,k}(t)\bold{g}^C_{n,l}(t)\sqrt{p_l(t)}s_l(t) \nonumber\\
&+\sum_{n=1}^{N} \bold{v}^H_{n,k}(t)\bold{z}_n(t),
\end{align}
where $\bold{v}_{n,k}(t) \in \mathbb{C}^{M \times 1}$ denotes the uplink receiver beamforming vector of RRH $n$ for the Tx UE of D2D pair $k$, $\bold{g}^{C}_{n,k}(t)\in \mathbb{C}^{M \times 1}$ is the CSI vector from RRH $n$ to the Tx UE of D2D pair $k$,
$p_k(t)$ represents the transmit power of D2D pair $k$, $s_k(t)$ is the data symbol transmitted by D2D pair $k$ with zero mean and unit variance, and $\textbf{z}_n(t) \in \mathbb{C}^{M \times 1}$ is the additive white Gaussian noise (AWGN) vector at RRH $n$, which is distributed as $\bold{z}_n(t)\sim\mathcal{CN}(\bold{0},\sigma^2\bold{I}_M)$.

As for C-RAN mode, we assume that the transmission rate of downlinks is no less than that of uplinks. This assumption can be established because of higher transmit power of RRHs {\cite{I:bibi2}}.
Therefore, using Shannon¡¯s formula, the achievable transmission rate in the unit of bit/s/Hz of D2D pair $k$ operating in C-RAN mode is given by
\begin{align}
\label{UL rate}
R^{C}_k(t)=& \log_2 \left(1+\frac{\displaystyle {p_k(t)}\left|\bold{v}^H_k(t)\bold{g}^C_k(t)\right|^2}
{\displaystyle \sum_{l\neq k}^{K} {p_l(t)}\left|\bold{v}^H_k(t)\bold{g}^C_l(t)\right|^2
+\sigma^2\|\bold{v}_k(t)\|^2_2}\right),
\end{align}
where $\bold{v}_{k}(t) \in \mathbb{C}^{NM \times 1}$ is the network-wide beamforming vector for the Tx UE of D2D pair $k$, $\bold{g}^C_k(t) \in \mathbb{C}^{NM \times 1}$ denotes the CSI vector from all
RRHs to the Tx UE of D2D pair $k$. Here, we have $\bold{v}_k=\left[(\bold{v}_{1,k})^T, \cdot\cdot\cdot, (\bold{v}_{N,k})^T\right]^T$ and
$\bold{g}^C_k=\left[(\bold{g}^C_{1,k})^T, \cdot\cdot\cdot, (\bold{g}_{N,k})^T\right]^T$.

Similarly, at slot $t$, the received signal of the Rx UE of D2D pair $i$ operating in D2D mode can be written as follows:
\begin{align}
\label{rece_signal d2d}
y^{D}_i(t)=g^{D}_{i,i}(t)\sqrt{p_i(t)}s_i(t)+\sum_{j\neq i}^{K}g^{D}_{j,i}(t)\sqrt{p_j(t)}s_j(t)+\varphi_i(t),
\end{align}
where $g^{D}_{i,i}(t)$ represents the channel gain from the Tx UE of D2D pair $i$ to the Rx UE of D2D pair $i$, $\varphi_i(t)$ denotes the AWGN at the Rx UE of D2D pair $i$, i.e.,
$\varphi_i(t)\sim\mathcal{CN}(0,\sigma^2)$.

The achievable transmission rate of D2D pair $i$ operating in D2D mode is given by
\begin{equation}
\label{D2D rate}
R^{D}_i(t)=\log_2 \bigg(1+\frac{{p_i(t)}{|{g}^{D}_{i,i}(t)|}^2}
{\displaystyle\sum_{j\neq i}^{K}{p_j(t)}{|{g}^{D}_{j,i}(t)|}^2+\sigma^2}\bigg).
\end{equation}

For mode selection, the binary mode selection indicator of D2D pair $k$ at slot $t$ is defined as follows:
\begin{equation}
x_k(t)=\left\{ \begin{array}{ll}
0,&\textrm{C-RAN mode }\\
1,&\textrm{D2D mode}.
\end{array} \right. \nonumber
\end{equation}
\indent Therefore, a general expression of the achievable transmission rate of D2D pair $k$ can be rewritten as
\begin{equation}
R_k(t)=(1-x_k(t))R^{C}_k(t)+x_k(t)R^{D}_k(t).
\end{equation}

\subsection{Definitions of Network Stability and Average Throughput}
Note that it is difficult to make a precise analysis on the whole end-to-end transmit latency, i.e., including the processing delay, queueing delay, transmission delay, and propagation delay. Since the queueing theory framework can establish the relationship among the queueing delay, the arrival rate, and the transmission rate{\cite{I:delay1}},
this paper just focuses on the queueing delay{\cite{I:bibi5}}--{\cite{I:BnB1}}, {\cite{I:lagrange}}, {\cite{I:slot}}.
To clarify the aforesaid relationship and the system delay requirement, it is necessary to introduce the concepts of the traffic buffering queue and the stability of the network.

Suppose that individual traffic buffering queues are maintained for all D2D pairs. At slot $t$, let $Q_k(t)$ represent the data queue length of D2D pair $k$ and denote the amount of stochastic traffic arrivals as $A_k(t)$, which is i.i.d. over slots with mean $\mathbb{E}\{A_k(t)\}=\lambda_k$. Therefore, the data queue length $Q_k(t)$ evolves according to{\cite{II:R1}}
\begin{equation}
\label{queue}
Q_k(t+1)=\max[Q_k(t)-R_k(t),0]+A_k(t).
\end{equation}
\indent \textit{Definition 1}: A discrete time process $U(t)$ is mean rate stable{\cite{II:R1}} if
\begin{equation}
\lim_{t \to \infty}\frac{\mathbb{E}\left\{|U(t)|\right\}}{t}=0
\end{equation}
and a network is stable if all individual queues are stable.

\textit{Remark 1}:
Note that the average delay can be depicted by the average data queue length according to the Little¡¯s Theorem{\cite{II:R1}}.
Definition 1 implies that the exogenous arrived data can be transmitted within a finite delay if network stability is guaranteed.
Furthermore, when a network of data queues is stable, the achieved average throughput can be represented by the time-averaged transmission rate{\cite{II:R1}}.

Therefore, the average throughput of D2D pair $k$ is defined as
\begin{equation}
\overline{R_k}=\lim_{T \to \infty}\frac{1}{T}\sum_{t=0}^{T-1}\mathbb{E}\{R_k(t)\}.
\end{equation}

\subsection{Problem Formulation}

Denote $\bold{x}(t)=\left\{x_1(t), \cdot\cdot\cdot, x_K(t)\right\}^T$ as $K$-dimensional binary mode selection vector at slot $t$, Similarly, denote $\bold{V}(t)\in\mathbb{C}^{NM\times K}$ and $\bold{p}(t)$ as the uplink receiver beamforming matrix and the power control vector at slot $t$, respectively. Let $P^{I}_{Dmax}$ and $P_{max}$ denote the interference tolerance threshold of D2D pairs and the peak transmit power, respectively.

Meanwhile, the total accumulated transmission rates of D2D pairs served by RRH $n$ should satisfy the fronthaul capacity constraint at slot $t$, which can be expressed as follows:
\begin{align}
\sum_{k=1}^{K}\mathbbm{1}\left\{\|\bold{D}_n\bold{v}_k(t)\|^2_2\right\}R_k(t)\leq C_n, \forall  n \in \mathcal{N},
\end{align}
where $\bold{D}_n=\{\underbrace{\bold{0}_M, \cdot \cdot \cdot, \bold{0}_M}_{n-1}, \bold{I}_M, \bold{0}_M, \cdot \cdot \cdot, \bold{0}_M\} \in  \mathbb{R}^{M \times NM}$, $\mathbbm{1}\left\{\|\bold{D}_n\bold{v}_k(t)\|^2_2\right\}$ is an indicator function equal to 0 if $\|\bold{D}_n\bold{v}_k(t)\|^2_2=0$, and 1 otherwise. $C_n$ denotes the fronthaul capacity limitation of RRH $n$. Note that $\bold{v}_{n,k}(t)$ can be represented via $\bold{v}_k(t)$, i.e., $\bold{v}_{n,k}(t)=\bold{D}_n\bold{v}_k(t)$.

At slot $t$, we aim to maximize the overall average throughput in C-RANs with D2D via joint mode selection, uplink beamforming design, and power control subject to network stability, interference, and fronthaul capacity constraints, which can summarized by the following stochastic optimization problem.
\begin{align}
\label{org}
&\mathop{\max}_{\bold{x}(t), \bold{V}(t), \bold{p}(t)} \quad \sum_{ k=1}^{K}\overline{R_k} \nonumber\\
&  s.t.  \quad \text{C1:}\,\, Q_k(t) \text{ is mean rate stable}, \forall k \in \mathcal{K}, t, \nonumber\\
& \qquad \,\, \text{C2:}\,\, \sum_{k=1}^{K} x_k(t)p_k(t)\leq P^{I}_{Dmax}, \forall t,\nonumber\\
& \qquad \,\, \text{C3:}\,\, x_k(t)\in\{0,1\}, \forall k\in \mathcal{K}, t, \nonumber\\
& \qquad \,\, \text{C4:}\,\, 0\leq p_k(t)\leq P_{max}, \forall k \in\mathcal{K}, t, \nonumber\\
& \qquad \,\, \text{C5:}\,\, \sum_{k=1}^{K}\mathbbm{1}\left\{\|\bold{D}_n\bold{v}_k(t)\|^2_2\right\}R_k(t)\leq C_n, \forall  n \in \mathcal{N}, t.
\end{align}

In (\ref{org}), C1 is the network stability constraint to guarantee all exogenous
arrived data of D2D pairs can be transmitted from the buffer within a finite time, i.e., the delay requirement of all D2D pairs can be fulfilled. Thus, the overall average throughput will be maximized under considering the scheduling fairness with this constraint.
C2 is the total transmit power constraint
on all D2D pairs operating in D2D mode in order to restrain both the inter-tier interference to D2D pairs operating in C-RAN mode and the intra-tier interference to D2D pairs operating in D2D mode. By appropriately setting interference tolerance threshold $P^{I}_{Dmax}$, the network operators can control both the inter-tier interference and the intra-tier interference caused by D2D pairs operating in D2D mode{\cite{I:c21}}. C3 indicates that any D2D pair can only operate in either C-RAN mode or D2D mode at slot $t$. C4 is the peak transmit power constraint. C5 is the fronthaul capacity constraint of RRH $n$.

Intuitively, the objective function of the stochastic optimization problem (\ref{org}) is the long-term time average of the expected transmission rate in each slot. C1 is the constraint on time-averaged variables. The binary mode selection variables, the mixed discrete and continuous fronthaul constraint C5 make the optimization problem NP-hard.
Theoretically, the optimal solution to (\ref{org}) can be obtained via dynamic programming techniques if the full statistical knowledge of both the time-varying channel conditions and the traffic arrivals are known.
However, it is challenging to get the statistics and is highly costly to calculate the optimal solution due to the curse of dimensionality{\cite{I:delay1}}, {\cite{I:delay2}}.
To this end, we resort to the Lyapunov optimization approach to design a cross-layer resource allocation algorithm, which
makes the online control policies at the beginning of each slot solely based on current CSI and QSI without requiring any priori knowledge of the stochastic processes.

\section{Overall Average Throughput Maximization}
In this section, based on the Lyapunov optimization technique, the stochastic optimization problem (\ref{org}) is transformed into a delay-aware resource allocation optimization problem. This non-convex mixed-integer nonlinear programming problem could be decomposed into three separate subproblems with respect to mode selection, uplink beamforming design, and power control, respectively. Finally, these three separate subproblems would be iteratively solved by the proposed JMSRA algorithm.

\subsection{Lyapunov Optimization}
The Lyapunov optimization framework has been proved to be particularly efficient and effective to optimize the
time average of the objective function subject to additional time-averaged constraints{\cite{II:R1}}, as the original stochastic optimization problem can be transformed into an instantaneous static optimization problem.
Thus, the classical drift-plus-penalty algorithm developed by the Lyapunov optimization technique can be directly exploited to tackle the stochastic optimization problem (\ref{org}).
In what follows, the definition of Lyapunov function and the Lyapunov drift is provided, both of which are used to derive the drift-plus-penalty expression.

Denote $\bold{\Theta}(t)=\left\{Q_k(t)|k\in\mathcal{K}\right\}$ as the vector of underlying data queues. According to{\cite{II:R1}}, the quadratic Lyapunov function is constructed as a scalar metric of queue congestion:
\begin{equation}
L(\bold{\Theta}(t))\triangleq \frac{1}{2}\sum_{k \in \mathcal{K}}{Q_k(t)}^2.
\end{equation}
\indent The one-slot conditional Lyapunov drift is introduced to push the Lyapunov function to a lower congestion state so that the network stability can be guaranteed, which is defined as
\begin{equation}
\vartriangle(\bold{\Theta}(t))=\mathbb{E}\left\{L(\bold{\Theta}(t+1))-L(\bold{\Theta}(t))|\bold{\Theta}(t)\right\}.
\end{equation}
\indent In addition, the drift-plus-penalty expression of (\ref{org}) is given by
\begin{equation}
\label{dpp}
\vartriangle(\bold{\Theta}(t))-V\mathbb{E}\left\{\sum_{k\in \mathcal{K}}{R_k}(t)|\bold{\Theta}(t)\right\},
\end{equation}
where the non-negative control parameter $V$ represents the importance weight placed on overall average throughput maximization, which can be adjusted by the network operators according to the performance requirement. More specifically, $V$ is a tuning parameter to control the performance gap between the
proposed algorithm and the optimal solution. With a larger $V$, the overall throughput can be closer to its optimum while incurring
a linearly increasing average delay.

The following lemma, proved in Appendix A, provides an upper bound of the drift-plus-penalty expression.

\textit{Lemma 1:} At any slot $t$, for any observed CSI and QSI, all parameters $V\geq 0$, all possible values of $\bold{\Theta}(t)$, the drift-plus-penalty expression (\ref{dpp}) satisfies the following inequality under any joint mode selection and resource allocation algorithms for C-RANs with D2D:
\begin{align}
\label{RHS}
\vartriangle & (\bold{\Theta}(t)) -V\mathbb{E}\left\{\sum_{k\in \mathcal{K}}{R_k}(t)|\bold{\Theta}(t)\right\}\nonumber\\
& \leq B+\sum_{k\in\mathcal{K}}{Q_k}(t)\mathbb{E}\{{A_k}(t)-{R_k}(t)|\bold{\Theta}(t)\}\nonumber\\
& \quad -V\mathbb{E}\left\{\sum_{k\in \mathcal{K}}{R_k(t)}|\bold{\Theta}(t)\right\},
\end{align}
where $B$ is a positive constant that satisfies
\begin{equation}
B\geq \frac{1}{2}\sum_{k\in \mathcal{K}}\mathbb{E}\left\{{R_k}(t)^2+{A_k}(t)^2|\bold{\Theta}(t)\right\}.
\end{equation}

According to the theory of Lyapunov optimization approach in{\cite{II:R1}},
rather than pushing the drift-plus-penalty expression (\ref{dpp}) to the minimum directly,
it is necessary to minimize the right-hand-side, i.e., the
upper bound of the drift-plus-penalty expression, of the inequality (\ref{RHS})
subject to the same constraints except the network stability constraint C1. Moreover,
with the help of the principle of opportunistically minimizing an expectation in{\cite{II:R1}}, the stochastic optimization problem (\ref{org}) can be transformed into the following delay-aware joint mode selection and resource allocation optimization problem:
\begin{align}
\label{ET}
&\mathop{\min}_{\bold{x}(t), \bold{V}(t), \bold{p}(t)} \quad  \sum_{k\in\mathcal{K}}Y_k(t)((1-x_k(t))R^{C}_k(t)+x_k(t)R^{D}_k(t))\nonumber\\
&\qquad s.t. \qquad\quad \text{C2}, \text{C3}, \text{C4}, \text{C5},
\end{align}
where $Y_k(t)$$=$$-(Q_k(t)+V)$, which can be easily calculated by the observed QSI at slot $t$.

It can be observed that the optimization problem (\ref{ET}) consists of
two layers. The external layer is the selection process of transmission mode, which involves the 0--1 integer optimization problem. The internal layer could be further split into two separate subproblems. The first problem is uplink beamforming design and the other is power control. These two layers could be decoupled and solved iteratively.

\subsection{Modified Branch and Bound Algorithm}
An effective way to solve the 0--1 integer optimization problem is the branch and bound method{\cite{I:BnB1}}. 
However, the computational complexity of branch and bound is $O(2^K)$ since a search for a complete $K$-order binary tree is required in the worst case.
Thus, the number of iterations increases exponentially with the number of D2D pairs, which makes it difficult to apply in practice. To further reduce the computational complexity, a modified version of the branch and bound method is proposed to efficiently solve the problematic mode selection problem in the following pages. For simplicity, the time slot index $t$ is dropped in the rest of this page.

Firstly, all binary mode selection indicators are relaxed to real continuous region $[0, 1]$. The corresponding root problem $Q_0$ can be reformulated as following:
\begin{align}
\label{BB}
Q_0\qquad &\mathop{\min}_{\bold{x}, \bold{V}, \bold{p}} \quad  \sum_{k\in\mathcal{K}}Y_k((1-x_k)R^{C}_k+x_kR^{D}_k)\nonumber\\
&\,\,\, s.t. \quad\, \text{C2}, \text{C4}, \text{C5},\nonumber\\
& \qquad \quad \,\, 0\leq x_k \leq1, \forall k \in \mathcal{K}.
\end{align}

Secondly, solve the relaxed problem and get the corresponding solution. Consider $\bold{x}^{\ast}_0$ and $Q^{\ast}_0$ as the optimal solution to $Q_0$ and the optimal value of the objective function of $Q_0$, respectively. If each element of $\bold{x}^{\ast}_0$ is integer, the solution obtained is the optimal solution and output $\bold{x}^{\ast}_0$. If not, the branching strategy is applied to $Q_0$. At each iteration, we branch a parent problem into two new subproblems. Differing from the conventional branch and bound method, two important points in the process of branching of the proposed modified branch and bound algorithm need to be clarified. First, we choose the non-integer element $x_{k^{'}}$ to be the branching variable. Specifically, $k^{'}$ is decided by
\begin{align}
\label{split}
k^{'}=\text{arg} \mathop{\max}_{k}\,\, \{R^{C}_k, R^{D}_k\}.
\end{align}

It is implied form (\ref{split}) that we choose the maximum partial derivative of the objective function of $Q_0$ with respect to $x_k (\forall k \in \mathcal{K})$ as the branching variable. This is because a larger partial derivative leads to more rapid convergence, resulting in the reduction of operation quantity.

Then, along with $x_{k^{'}}$, we can split $Q_0$ into two new sub-problems $Q_1$ and $Q_2$. These two new formed sub-problems can be generally expressed as
\begin{align}
Q_1\qquad &\mathop{\min}_{\bold{x}, \bold{V}, \bold{p}} \quad  \sum_{k\in\mathcal{K}}Y_k((1-x_k)R^{C}_k+x_kR^{D}_k)\nonumber\\
&\,\,\, s.t. \quad\, \text{C2}, \text{C4}, \text{C5},\nonumber\\
&\qquad \quad \,\, 0\leq x_k \leq1, \forall k \in \mathcal{K}\backslash\{{k^{'}}\},\nonumber\\
&\qquad \quad \,\, x_{k^{'}}=0,
\end{align}
\begin{align}
Q_2\qquad &\mathop{\min}_{\bold{x}, \bold{V}, \bold{p}} \quad  \sum_{k\in\mathcal{K}}Y_k((1-x_k)R^{C}_k+x_kR^{D}_k)\nonumber\\
&\,\,\, s.t. \quad\, \text{C2}, \text{C4}, \text{C5},\nonumber\\
& \qquad \quad \,\, 0\leq x_k \leq1, \forall k \in \mathcal{K}\backslash\{{k^{'}}\},\nonumber\\
&\qquad \quad \,\, x_{k^{'}}=1.
\end{align}

Second, the rule for branch strategy is based on depth first strategy, which means that we select two problems with the
smallest lower bound to branch until it reaches a binary solution or reaches infeasibility according to the fact that the optimal solution is most likely to be contained in it{\cite{I:BnB2}}. With these two modification, the modified branch and bound method utilized in this paper consists in restricting the search to 2 survival paths in the branch and bound tree. In the worst case, it can be deduced that the computational complexity of the modified branch and bound method is $O(2K)$, which grows linearly with the number of D2D pairs and is scalable for large-scale C-RANs with D2D.

The process of branching and bounding will be repeated until the optimal solution to the relaxed sub-problem satisfies all
integer constraint with minimum value of the objective function. By branching, we can obtain better and better solution. After mode selection, $x_k$ ($\forall k\in\mathcal{K}$) can be removed, the optimization problem (\ref{ET}) is simplified as
\begin{align}
\label{ETII}
\mathop{\min}_{\bold{V}, \bold{p}} \quad  &\sum_{k\in\mathcal{C}}Y_kR^{C}_k+\sum_{i\in\mathcal{J}}Y_iR^{D}_i \nonumber\\
s.t. \quad & \sum_{i \in \mathcal{J}} p_i \leq P^{I}_{Dmax},  \nonumber\\
& 0 \leq p_i \leq P_{max}, \forall i \in \mathcal{C}\cup\mathcal{J}, \nonumber\\
& \sum_{ k\in \mathcal{C}}\mathbbm{1}\left\{\|\bold{D}_n\bold{v}_k\|^2_2\right\}R_k\leq C_n, \forall  n \in \mathcal{N},
\end{align}
where $\mathcal{C}$ and $\mathcal{J}$ denote the sets of D2D pairs operating in C-RAN mode and D2D mode, respectively. Here, we have $|\mathcal{C}|+|\mathcal{J}|=K$.

\subsection{Uplink Beamforming Design Algorithm}

If the power control results of all Tx UEs of D2D pairs operating in D2D mode are given, the separate uplink beamforming design problem can be reformulated as
\begin{align}
\label{norm}
\mathop{\min}_{\bold{V}, \bold{p}} \quad  &\sum_{k\in\mathcal{C}}Y_kR^{C}_k  \nonumber\\
s.t. \quad & 0 \leq p_k \leq P_{max}, \forall k \in \mathcal{C}, \nonumber\\
     \quad & \sum_{ k\in \mathcal{C}}\mathbbm{1}\left\{\|\bold{D}_n\bold{v}_k\|^2_2\right\}R_k\leq C_n, \forall  n \in \mathcal{N}.
\end{align}

The indicator function in (\ref{norm}) can be equivalently expressed as an ${\ell}_0$-norm of a scalar, which is the number of nonzero entries in a vector. This equivalent expression allows us approximately transform a nonconvex ${\ell}_0$-norm optimization objective into a convex reweighted ${\ell}_1$-norm{\cite{I:Fronthaul}}. Therefore, the indicator function $\mathbbm{1}\left\{\|\bold{D}_n\bold{v}_k(t)\|^2_2\right\}$ can be written as follows:
\begin{align}
\mathbbm{1}\left\{\|\bold{D}_n\bold{v}_k\|^2_2\right\}={\left\|\|\bold{D}_n\bold{v}_k\|^2_2\right\|}_0.
\end{align}

The fronthaul capacity constraint in (\ref{norm}) can be reformulated as
\begin{align}
\sum_{ k\in \mathcal{C}}{\beta}_{n,k}{\|\bold{D}_n\bold{v}_k\|}^2_2R_k\leq C_n, \forall  n \in \mathcal{N}.
\end{align}
where ${\beta}_{n,k}$ is a constant weight and is updated iteratively according to
\begin{align}
{\beta}_{n,k}=\frac{1}{{\|\bold{D}_n\bold{v}_k\|}^2_2+\tau}, \forall k \in \mathcal{C}, n \in \mathcal{N}
\end{align}
with a small constant regularization factor $\tau >0$ and $\bold{v}_k$ from the previous iteration.

Because of the fact that the transmission rate $R_k$ is related to both the objective function and the
constraints, the optimization problem (\ref{norm}) consists of the fronthaul capacity constraint is still difficult to deal with even with the adoption of the above approximation. To address this difficulty, an iterative scheme with the fixed transmission rate ${\hat{R}}_k$ obtained from
the previous iteration is used. Thus, the optimization problem now can be rewritten as
\begin{align}
\label{BF}
\mathop{\min}_{\bold{V}, \bold{p}} \quad  &\sum_{k\in\mathcal{C}}Y_kR^{C}_k  \nonumber\\
s.t. \quad & 0 \leq p_k \leq P_{max}, \forall k \in \mathcal{C}, \nonumber\\
     \quad & \sum_{ k\in \mathcal{C}}{\beta}_{n,k}{\|\bold{D}_n\bold{v}_k\|}^2_2{\hat{R}}_k\leq C_n, \forall  n \in \mathcal{N}.
\end{align}

Obviously, the optimization problem (\ref{BF}) is still non-convex, which is difficult to be solved directly. Fortunately, inspired by the celebrated duality theory for uplink and downlink beamforming{\cite{I:ULDL}}, the equivalence between the weighted sum rate maximization problem and the penalized WMMSE problem for multiple-input and multiple-output interfering channel{\cite{I:WMMSE2}}, the problem (\ref{BF}) has the same optimal solution as the following WMMSE minimization problem:
\begin{align}
\label{wmmse}
&\mathop{\min}_{\bold{w}_{k}, \rho_k, \mu_k,} \quad  \sum_{k\in\mathcal{C}}Y^{'}_k(\rho_ke_k-\log\rho_k)  \nonumber\\
&\quad\, s.t. \quad \quad \text{C6:}\,\,{\|\bold{w}_{k}\|}^2_2 \leq P_{max}, \forall k \in \mathcal{C}, \nonumber\\
&\quad\,\, \quad\qquad\,\text{C7:}\,\,\sum_{ k\in \mathcal{C}}{\beta}_{n,k}\|\bold{D}_n\bold{w}_k\|^2\frac{{\hat{R}}_k}{{\hat{p}}_k}\leq C_n, \forall  n \in \mathcal{N},
\end{align}
where $Y^{'}_k=Q_k+V$, $\bold{w}_{k} \in \mathbb{C}^{NM \times 1}$ is a virtual network-wide downlink transmit beamformer for the Tx UE of D2D pair $k$ to jointly solve the uplink receiver beamforming and power control problem, ${\hat{p}}_k$ is the transmit power of D2D pair $k$ operating in C-RAN mode obtained from
the previous iteration, $\rho_k$ denotes a positive mean-square estimation (MSE) weight, and $e_k$ is the corresponding MSE error.

Under the MMSE receiver $\mu_k\in\mathbb{C}$, $e_k$ is defined as
\begin{align}
\label{ek}
e_k&= \mathbb{E}\left\{{(\mu^H_ky^{C}_k-s_k)}^2\right\} \nonumber\\
   &= \mu^H_k\left(\sum_{l\in\mathcal{C}\cup\mathcal{J}}
{(\bold{g}^{C}_{l})}^H\bold{w}_k\bold{w}^H_k\bold{g}^{C}_{l}+\frac{\sigma^2}{{\hat{p}}_k}\right)\mu_k\nonumber\\
   &\quad-2\mathrm{Re}\big\{\mu^H_k{(\bold{g}^{C}_{k})}^H\bold{w}_k\big\}+1.
\end{align}

Problem (\ref{wmmse}) is convex with respect to each of the individual optimization variables when fixing the others.
Therefore, problem (\ref{wmmse}) can be solved efficiently by iterating through $\rho_k$, $\mu_k$, and $\bold{w}_{k}$ with block coordinate descent method{\cite{I:WMMSE2}}.

The optimal MSE weight $\rho_k$ under fixed $\mu_k$ and $\bold{w}_{k}$ is given by
\begin{equation}
\label{rhok}
\rho_k=e^{-1}_k, \forall k \in \mathcal{C}.
\end{equation}

The optimal receiver $\mu_k$ under fixed $\rho_k$ and $\bold{w}_{k}$ can be derived as
\begin{align}
\label{muk}
\mu_k=\frac{{(\bold{g}^{C}_{k})}^H\bold{w}_k}{\displaystyle\sum_{l\in\mathcal{C}\cup\mathcal{J}}
   {(\bold{g}^{C}_{l})}^H\bold{w}_k\bold{w}^H_k\bold{g}^{C}_{l}+\frac{\sigma^2}{{\hat{p}}_k}},  \forall k \in \mathcal{C}.
\end{align}

Under fixed $\rho_k$ and $\mu_k$, the optimization problem to find the optimal transmit beamformer $\bold{w}_{k}$ can be expressed as follows:
\begin{align}
\label{QCQP}
\mathop{\min}_{\bold{w}_{k}} \quad  &\sum_{k\in\mathcal{C}\cup\mathcal{J}}\bold{w}^H_{k}
\left(\sum_{l\in\mathcal{C}\cup\mathcal{J}}Y^{'}_l\rho_l\mu^H_l\bold{g}^{C}_{l}(\bold{g}^{C}_{l})^H\mu_l\right)\bold{w}_{k}  \nonumber\\ &-2\sum_{k\in\mathcal{C}\cup\mathcal{J}}Y^{'}_k\rho_k\mathrm{Re}\left\{\mu^H_k{(\bold{g}^{C}_{k})}^H\bold{w}_k\right\} \nonumber\\
s.t. \quad & \text{C6}, \text{C7}.
\end{align}

The optimization problem (\ref{QCQP}) is a quadratically constrained quadratic programming (QCQP) problem, which can be solved
via a standard convex optimization solver such as Matlab software for disciplined
convex programming (CVX) \cite{II:CVX}.

\subsection{Power Control Algorithm}

According to the duality theory in \cite{I:ULDL}, we have that $\bold{v}^{\ast}_k=[(\bold{v}^{\ast}_{1,k})^T, \cdot\cdot\cdot, (\bold{v}^{\ast}_{n,k})^T]^T=\bold{w}^{\ast}_k$, where
$\bold{v}^{\ast}_k$ and $\bold{w}^{\ast}_{k}$ are the optimal solutions to the problems (\ref{BF}) and (\ref{wmmse}), respectively.
With $\bold{v}_k=\bold{w}^{\ast}_k$, the optimization problem (\ref{BF}) becomes a power control optimization problem for sum-rate maximization, which can be rewritten as follows:
\begin{align}
\label{powercran}
\mathop{\min}_{\bold{p}} \quad  &\sum_{k\in\mathcal{C}}Y_k{R^{C}_k}^{\ast}  \nonumber\\
s.t. \quad & 0 \leq p_k \leq P_{max}, \forall k \in \mathcal{C},
\end{align}
where ${R^{C}_k}^{\ast}= \log_2 \left(1+\frac{\displaystyle {p_k}\left|(\bold{w}^{\ast}_k)^H\bold{g}^C_k\right|^2}
 {\sum_{l\neq k}^{K} {p_l(t)}\left|(\bold{w}^{\ast}_k)^H\bold{g}^C_l\right|^2+\sigma^2\|\bold{w}^{\ast}_k\|^2_2}\right)$.

Furthermore, when the power control results of all Tx UEs of D2D pairs operating in C-RAN mode are given, the power control problem of the Tx UEs of D2D pairs operating in D2D mode can be reformulated as
\begin{align}
\label{powerd2d}
\mathop{\min}_{\bold{p}} \quad  &\sum_{i\in\mathcal{J}}Y_iR^{D}_i  \nonumber\\
s.t. \quad & \sum_{i\in\mathcal{J}}p_i \leq P^{I}_{Dmax},          \nonumber\\
           & 0 \leq p_i \leq P_{max}.
\end{align}

According to the optimal power control solution in \cite{I:Power1}, it can be proven that both optimization problem (\ref{powercran}) and (\ref{powercran}) are convex
with respect to any one of the optimization variable $p_k (k \in \mathcal {C})$ or $p_i (i \in \mathcal {J})$ when the other variables $p_l (l \in \mathcal {C}, l\neq k)$ or $p_j (j \in \mathcal {J}, j\neq i)$ are fixed. Both problem can be solved iteratively.

Due to the space limitation, we take (\ref{powerd2d}) as an example,
and the associated Lagrangian function of the optimization problem (\ref{powerd2d}) subject to interference and transmit power constraints is given by
\begin{align}
\mathrm{L}(\bold{p}, \delta, \boldsymbol{\omega})=&\sum_{i\in\mathcal{J}}Y^{'}_iR^{D}_i+\delta\big(P^{I}_{Dmax}-\sum_{i\in\mathcal{J}}p_i\big)\nonumber\\
&+\sum_{i\in\mathcal{J}}{\omega}_i(P_{max}-p_i),
\end{align}
where $\boldsymbol{\omega}=[{\omega}_1, \cdots, {\omega}_{\left|\mathcal{J}\right|}]$ and $\delta$ are non-negative Lagrangian multiplier vector and multiplier, respectively. These Lagrange multipliers can be updated by using the gradient method {\cite{I:lagrange}}.

The gradient of $\mathrm{L}(\bold{p}, \delta, \boldsymbol{\omega})$ with respect to $p_i$ ($\forall i\in\mathcal{J}$) should be equal to zero. Therefore, the optimal power control of Tx UE of D2D pair $i$ can be obtained by
\begin{equation}
p_i=\frac{Y^{'}_i}{({\omega}_i+\delta)\ln2}-\frac{\sum_{j\in \mathcal{C}\cup \mathcal{J},j\neq i}{p_j}{|{g}^{D}_{j,i}|}^2+\sigma^2}{{|{g}^{D}_{i,i}|}^2}.
\end{equation}

Furthermore, the optimal transmit power $p^{\ast}_i$ of the Tx UE of D2D pair $i$ can be determined as follows:
\begin{equation}
\label{poweropt}
p^{\ast}_i=\max[0, p_i].
\end{equation}

Finally, the mode selection vector, network-wide beamforming vector, and power control vector would be iteratively computed by the BBU pool until their solutions converge.
The main steps of the proposed JMSRA algorithm 
are summarized in Algorithm 1.
\begin{algorithm}[!htp]
\caption{JMSRA Algorithm at Slot $t$}
\begin{algorithmic}[1]
\STATE \textbf{Initialize} all primal variables;
\STATE Select tolerance $\triangle > 0$ and iteration number $n=1$;
\REPEAT
\STATE \textbf{Initialize} the problem list with the root problem $Q_0$ and set its upper bound as $UB(Q_0)=\infty$;
\STATE \textbf{While} the problem list is not empty \textbf{Do}
\STATE Select the problem from the problem list that has the smallest lower bound by applying bounding strategy. Obtain its optimal mode selection solution $\bold{x}^{\ast}_{k}$ and lower bound $LB(Q_k)$, where $k$ is the node number;
\STATE \textbf{If} $\bold{x}^{\ast}_k$ is infeasible or $LB(Q_k)>UB(Q_0)$, discard the problem. \textbf{Else if}
        all elements in $\bold{x}^{\ast}_k$ are integers and $LB(Q_k)<UB(Q_0)$, set $\bold{x}^{\ast}=\bold{x}^{\ast}_k$, $UB(Q_0)=LB(Q_k)$ and then discard the problem.
       \textbf{Otherwise}, branch the problem into two new sub-problems along the determinate split index and add these new sub-problems to the problem list;
\STATE \textbf{End while}
\REPEAT
\STATE Fix $\bold{w}_{k} \left(\forall k\in\mathcal{C}\right)$, compute the MSE $e_k$ and the MMSE receiver $\mu_k$ according to (\ref{ek}) and (\ref{muk});
\STATE Update the MSE weight $\rho_k$ according to (\ref{rhok});
\STATE Find the optimal downlink transmit beamformer $\bold{w}^{\ast}_{k}$ under fixed $\mu_k$ and $\rho_k$ by solving problem (\ref{QCQP});
\STATE Set $\bold{v}^{\ast}_k=\bold{w}^{\ast}_k$, then obtain the optimal transmit power of the Tx UE of D2D pair $k$ operating in C-RAN mode by solving problem (\ref{powercran});
\STATE Compute the optimal transmit power of the Tx UE of D2D pair $i$ ($\forall i\in\mathcal{J}$) operating in D2D mode according to (\ref{poweropt});
\STATE Update $\bold{w}_k$, ${\beta}_{n,k}$, ${\hat{R}}_k$, $p_k$, $Y_k$, and $p_i$;
\UNTIL convergence
\STATE Compute the achievable transmission rate $R^{(n)}_i$;
\STATE $n=n+1$;
\UNTIL{$\left|R^{(n+1)}_i-R^{(n)}_i\right|\leq \triangle $}
\STATE \textbf{Return} the optimal solution as $\{\bold{x}^{\ast},\bold{V}^{\ast}, \bold{p}^{\ast}\}$;
\end{algorithmic}
\end{algorithm}

For the algorithm given above, although a rigorous theoretical proof of convergence is not available, simulation results in next section show that it can converge quickly in approximately 20 iterations under a appropriate initialization. The proposed JMSRA algorithm based on the modified branch and bound method and the WMMSE approach iteratively solve four subproblems, i.e., problem (\ref{BB}), (\ref{QCQP}), (\ref{powercran}), and (\ref{powerd2d}). Since the power control optimization problems (\ref{powercran}) and (\ref{powerd2d}) are convex, which can be efficiently solved by the fixed-point algorithm{\cite{I:Power2}}, the computational complexity of the proposed JMSRA algorithm is dominated by solving the integer programming problem (\ref{BB}) and the QCQP problem (\ref{QCQP}). In the worst case, when all D2D pairs operate in C-RAN mode, there are total $NKM$ variables in the QCQP problem and the computation complexity of using interior-point method to solve this problem is approximately $O((NKM)^{3.5})$ \cite{I:Fronthaul}. Therefore, the overall complexity of the proposed JMSRA algorithm is $O(2K(NKM)^{3.5})$.

\section{Numerical Results}

In this paper, to simplify the simulation, one radio resource block is considered for all RRHs and D2D pairs. If more RRHs and D2D pairs are evaluated in practise, the resource block allocation schemes proposed in{\cite{I:rb1}}, {\cite{I:rb2}} can be directly used. We simulate the problem in a C-RANs with D2D system consisting of $N=3$ RRHs and $K=6$ D2D pairs. The D2D pairs are uniformly and independently distributed in the square area of 0.5 $\times$ 0.5 km$^2$. Each pair of D2D UEs are distributed within distance limit. We set that each RRH is configured with $M=2$ antennas. Besides, the pathloss models for C-RANs and D2D links are modeled as $128.1+37.6\log(d)${\cite{I:WMMSE2}} and $148+40\log(d)${\cite{I:bibi2}}, respectively, where $d$ is the propagation distance in kilometer. In each slot, the fast fading channel gain is generated as i.i.d. complex Gaussian random variable with unit variance{\cite{I:BnB1}}. The noise power spectrum density is given as $\sigma^2=-174$ dBm/Hz{\cite{I:bibi2}}.
We set $P^I_{Dmax}=29$ dBm and $P_{max}=23$ dBm {\cite{I:ulpower}}. The mean arrival rate $\lambda_i$ of the Poisson traffic arrivals is assumed to be the same for all D2D pairs {\cite{I:c21}}.
The slot in this paper is identical as the frame in LTE systems and is set to be 10 milliseconds {\cite{I:slot}}, {\cite{I:ulpower}}. Note that a longer slot length has advantages of achieving better throughput performance gain with less signaling overhead.
To evaluate the performances of the proposed JMSRA algorithm, a Monte Carlo based system-level simulator has been built{\cite{I:slot}}. Each point of the simulation results is averaged over $5000$ slots.

The performance of our proposal is compared with
that of a C-RAN mode algorithm (C-RAN Mode) and a D2D mode algorithm (D2D Mode), described as follows:
\begin{itemize}
\item[\emph{$\bullet$}]
\textbf{C-RAN mode algorithm (C-RAN Mode)}: In this baseline, all D2D pairs choose to operate in C-RAN mode in each slot. A beamforming design scheme for downlink C-RANs proposed in{\cite{I:WMMSE2}} is utilized to solve the optimization problem (\ref{wmmse}), where the beamforming design algorithm is identical as that of JMSRA algorithm.
Since the simulation results of this algorithm could represent the performance of pure C-RAN scenarios, it can be regarded as a performance baseline for the proposed JMSRA algorithm.
\item[\emph{$\bullet$}]
\textbf{D2D mode algorithm (D2D Mode)}: In this baseline, all D2D pairs select D2D mode in each slot, and the overall average throughput is maximized by solving the power control subproblem. A power control algorithm based on Lagrange dual decomposition proposed in{\cite{I:Power1}} is used to find the optimal transmit power for the optimization problem (\ref{powerd2d}).
\end{itemize}

\begin{figure}[!htp]
\centering
\includegraphics[width=0.43\textwidth]{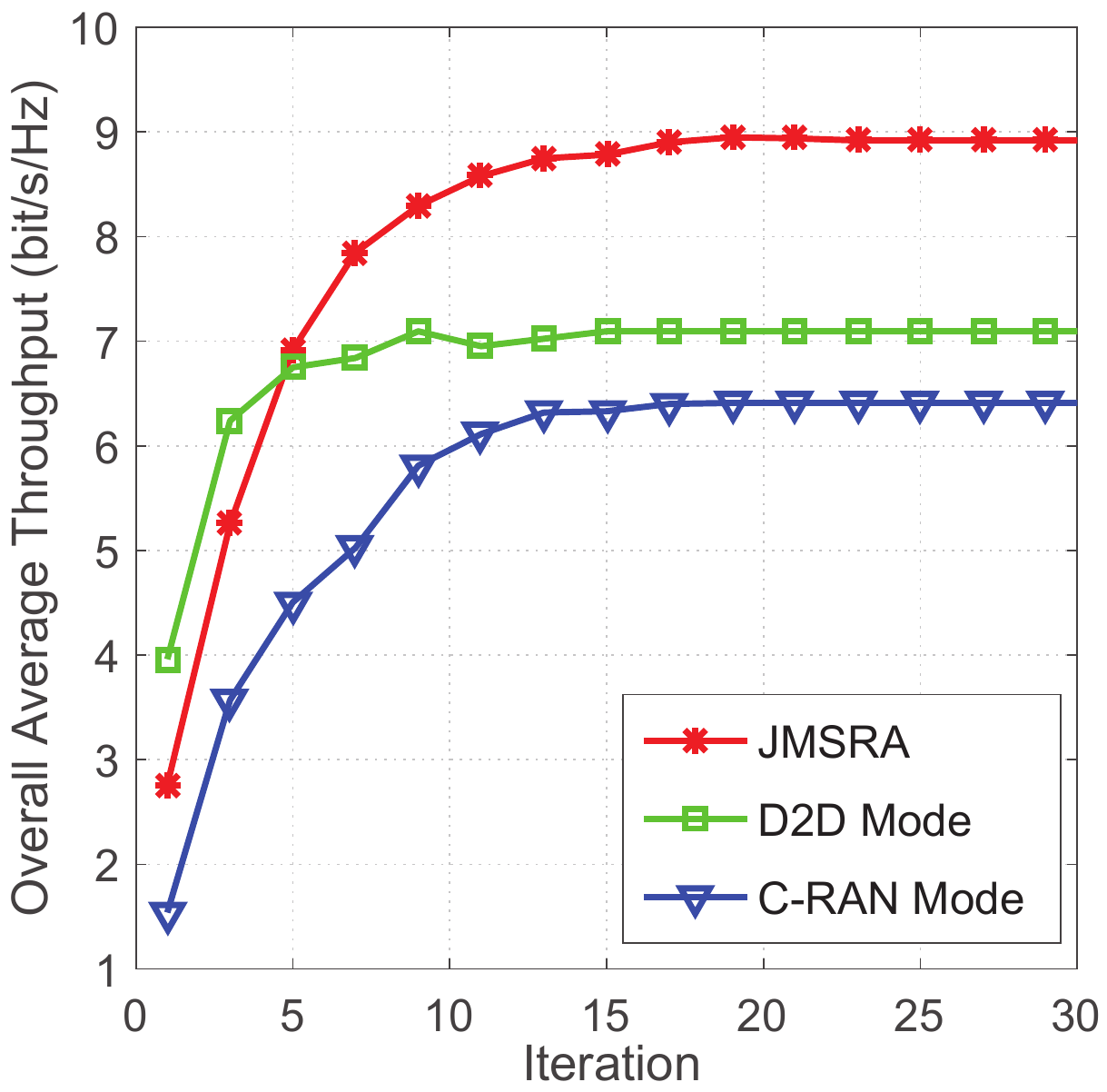}
\caption{Convergence behavior of different algorithms versus number of iterations.}\vspace*{-10pt}
\label{exp4}
\end{figure}

\indent Fig. \ref{exp4} presents the convergence behavior of three algorithms under the same initialization. It is observed that all three algorithms can always converge to stationary points in approximately 20 iterations. Moreover, a higher overall average throughput can be achieved by the JMSRA algorithm compared with both the C-RAN Mode and D2D Mode algorithms at the cost of lower speed of convergence.

\begin{figure}[!htp]
\centering
\includegraphics[width=0.43\textwidth]{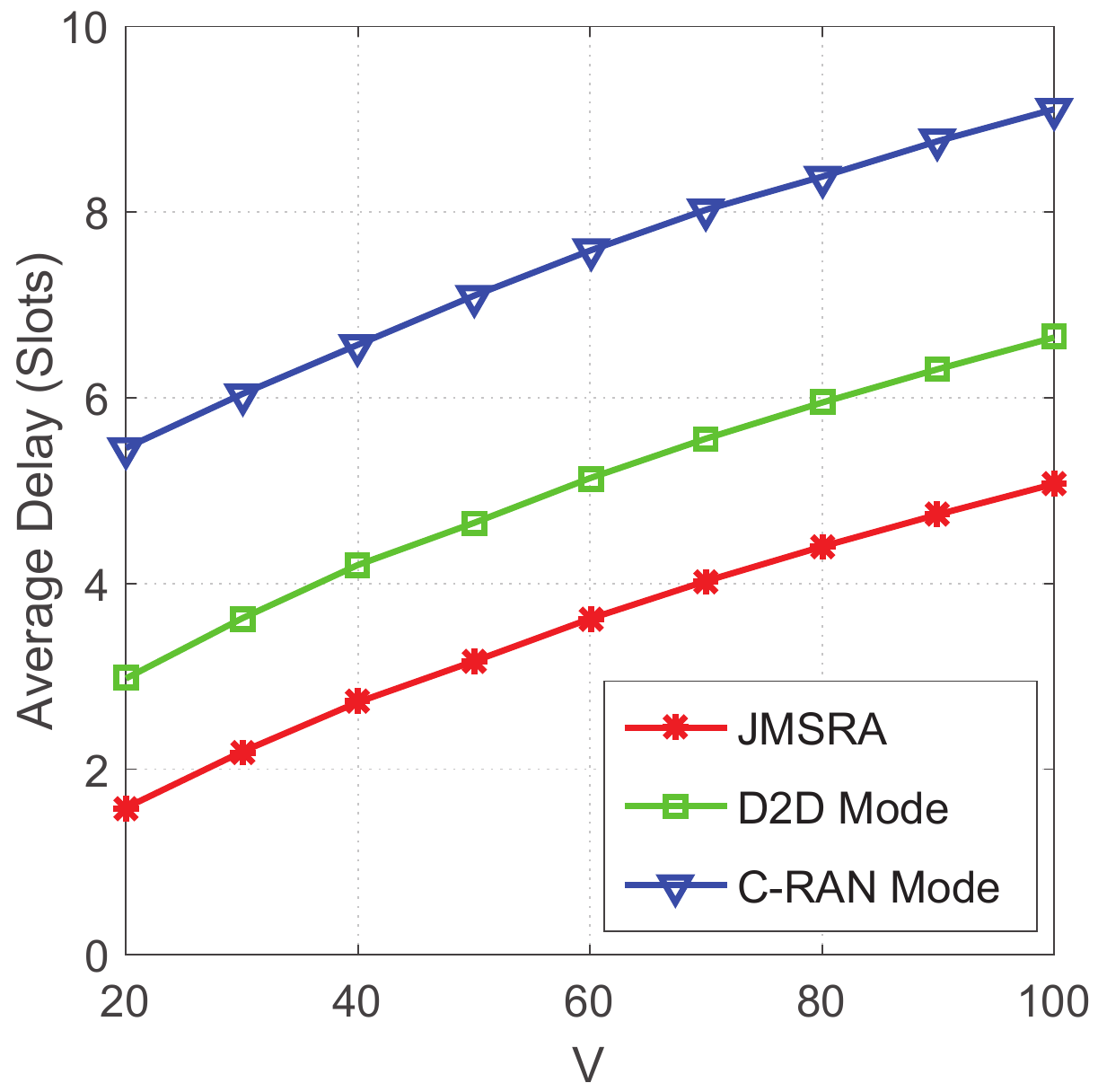}
\caption{Average delay versus control parameter $V$.}\vspace*{-10pt}
\label{exp1}
\end{figure}
\begin{figure}[!htp]
\centering
\includegraphics[width=0.43\textwidth]{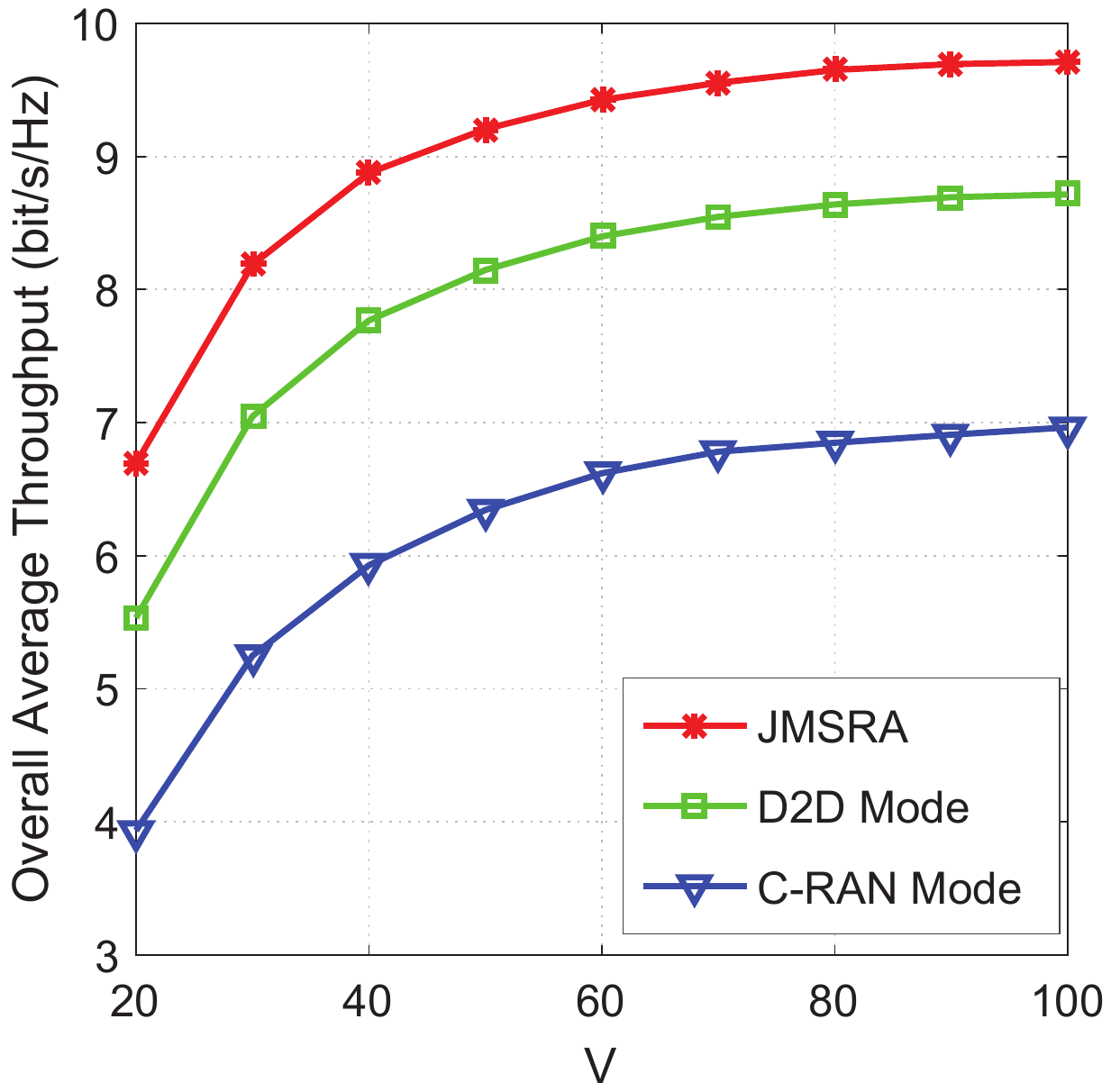}
\caption{Overall average throughput versus control parameter $V$.}
\label{exp2}\vspace*{-1em}
\end{figure}

\indent In Fig. \ref{exp1} and Fig. \ref{exp2}, we evaluate the average delay and the overall average throughput against the control parameter $V$, respectively. For all algorithms, the average delay, i.e., average queue length, shown in Fig. \ref{exp1} grows linearly at $O(V)$ under the given mean traffic arrival rate $\lambda=1$ bit/slot/Hz. A larger $V$ leads to higher average delay because of the fact that network systems with a larger $V$ emphasize less on delay performance.
The overall average throughput versus different control parameter $V$ is plotted in Fig. \ref{exp2}. The overall average throughput is increasing and convex in $V$ for all algorithms and increases toward the optimum at the speed of $1-O\left(\frac{1}{V}\right)$ as $V$ increases, which can be understood by the fact that greater emphasis is placed on the overall average throughput when $V$ increases.

\begin{figure}[!htp]
\centering
\includegraphics[width=0.43\textwidth]{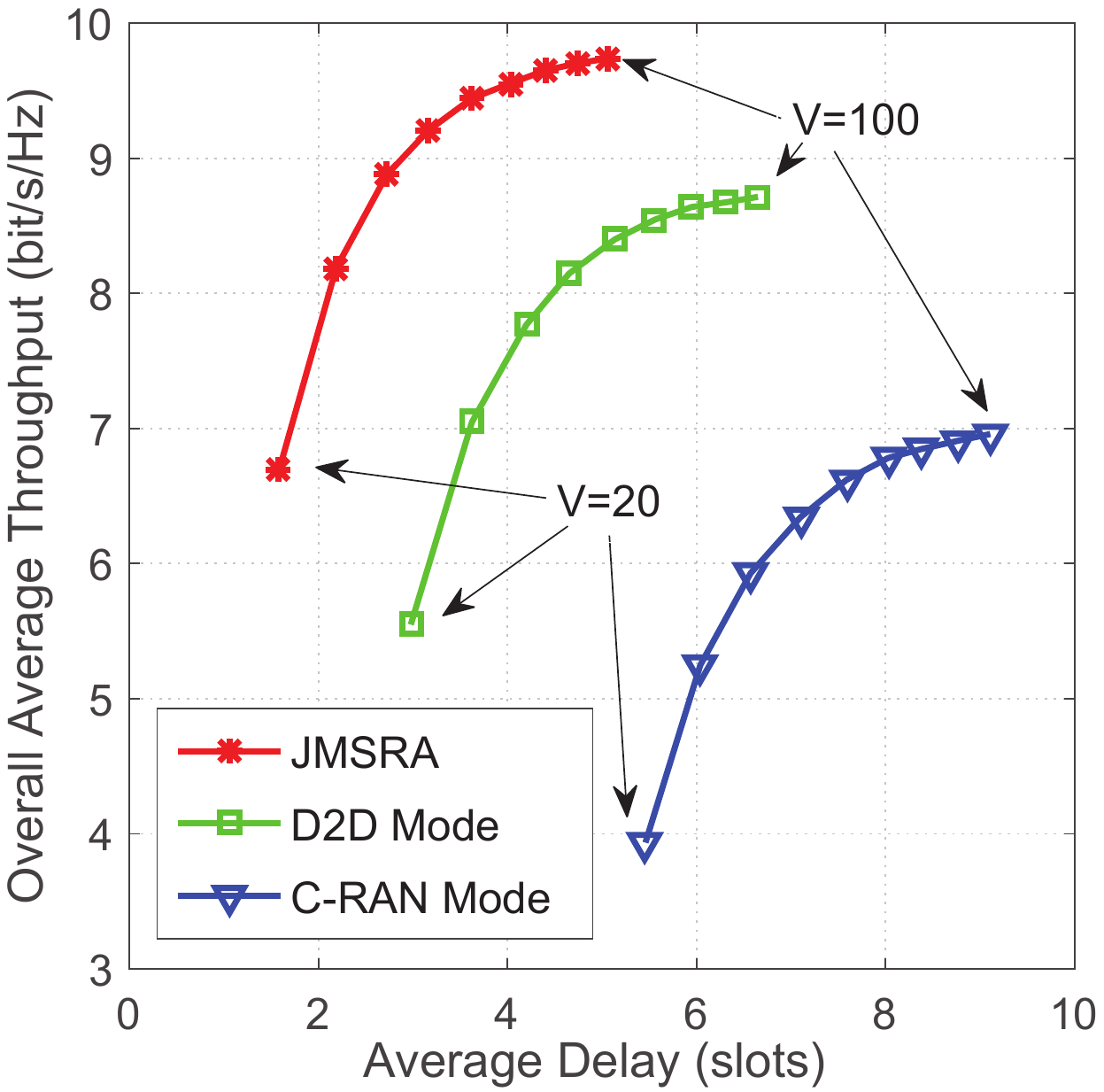}
\caption{Quantitative average throughput-delay tradeoff}
\label{exp6}\vspace*{-1em}
\end{figure}

\indent Combining with Fig. \ref{exp1} and Fig. \ref{exp2},
the JMSRA algorithm always provide better performance gains than the other two algorithms under arbitrary $V$ in terms of both average delay and overall average throughput.
The proposal always searches the best transmission mode for each D2D pair while the mode selection strategy is not available for the other two algorithms. D2D mode is preferable when the distance between a D2D pairs is small or far away from the RRHs. Thus, introducing D2D into C-RANs can improve the system throughput and reduce latency due to the reuse gain and the proximity gain from D2D communications.
The C-RAN mode is more beneficial when the distance between a D2D pair is large or the intra-tier interference among D2D pairs is severe since the BBU pool is more powerful in interference management.

Besides, the [$1-O\left(\frac{1}{V}\right),O(V)$] tradeoff relationship between overall average throughput and delay of various algorithms is illustrated in Fig. \ref{exp6} by varying the control parameter $V$.
Intuitively, when the average delay of the network system is small, slightly increasing $V$ can achieve a significant rising of the overall average throughput. On another hand, when the average delay is considerable high, decreasing $V$ will further proportionally improve the delay performance at the cost of only a very small amount of overall average throughput reduction. Moreover, the JMSRA algorithm provides significantly better average throughput-delay tradeoff and provides a simple approach to compromise the average throughput-delay performance on demand. Specifically, if network systems pursue for a better throughput performance, a larger $V$ is necessary. Otherwise, a smaller $V$ is preferable if the network system aims for a lower latency. In conclusion, the JMSRA algorithm provide a flexible and effective way to balance the average throughput-delay tradeoff, all we need to do is to select an appropriate control parameter $V$.

\begin{figure}[!htp]
\centering
\includegraphics[width=0.43\textwidth]{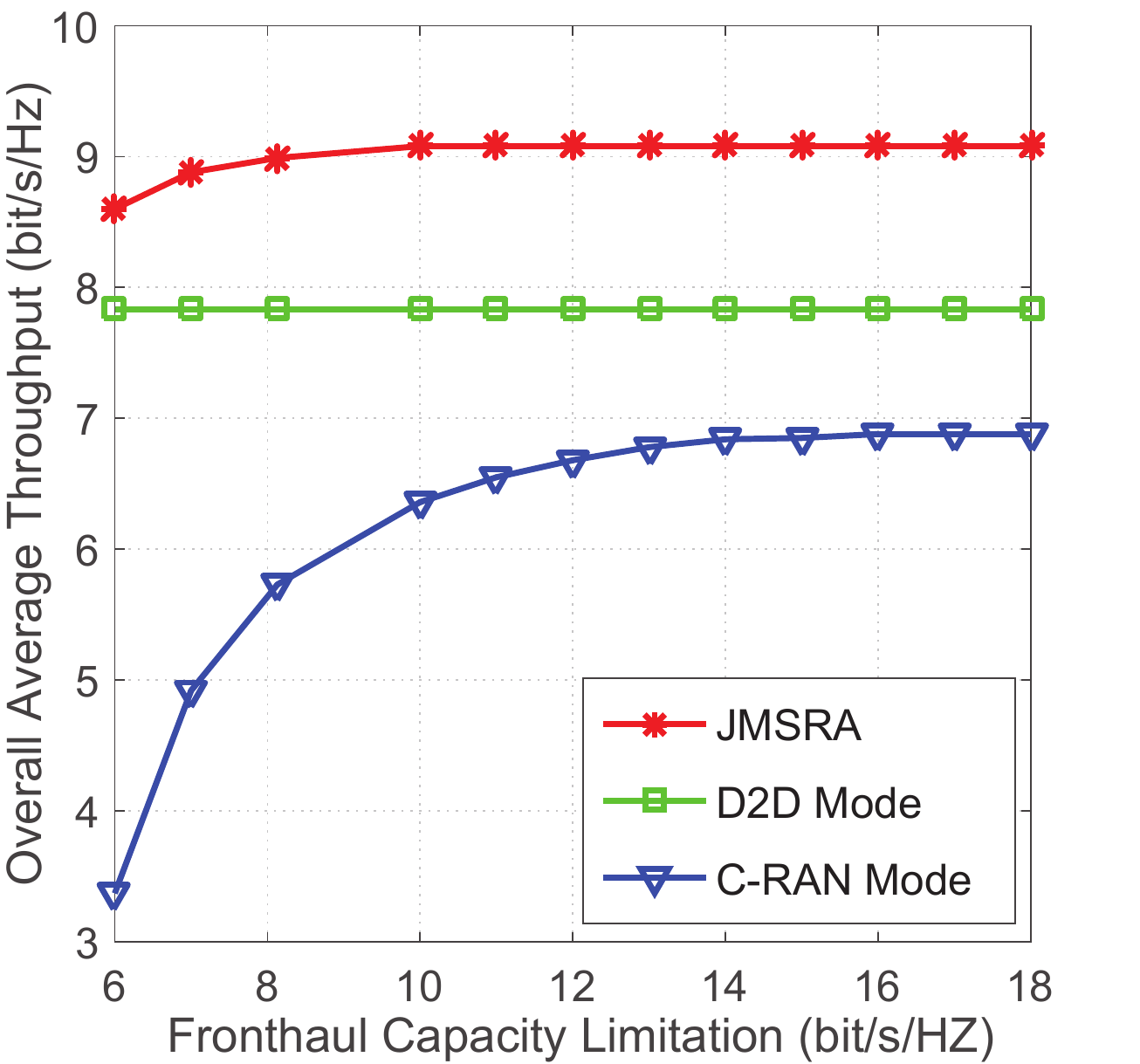}
\caption{Overall average throughput versus fronthaul capacity limitation.}
\label{exp3}\vspace*{-1em}
\end{figure}

Fig. \ref{exp3} shows the overall average throughput performance for all three algorithms under finite capacity of fronthaul. It can be seen that the proposed JMSRA algorithm achieves higher overall average throughput, which demonstrates the benefits of D2D deployment in C-RANs. Since D2D communications can offload traffic from C-RAN uplinks to D2D links, the heavy burdens of the constrained fronthaul can be alleviated. Therefore, higher overall average throughput and lower average delay can be achieved by the C-RAN uplinks. In addition, when the fronthaul capacity limitation increases, the performance gap between the JMSRA algorithm and the C-RAN Mode algorithm becomes smaller. The fact behind this is that the capacity constrained fronthaul links restrain the transmission rate, resulting in a significantly negative influence on SE performance of C-RANs.

\begin{figure}[!htp]
\centering
\includegraphics[width=0.43\textwidth]{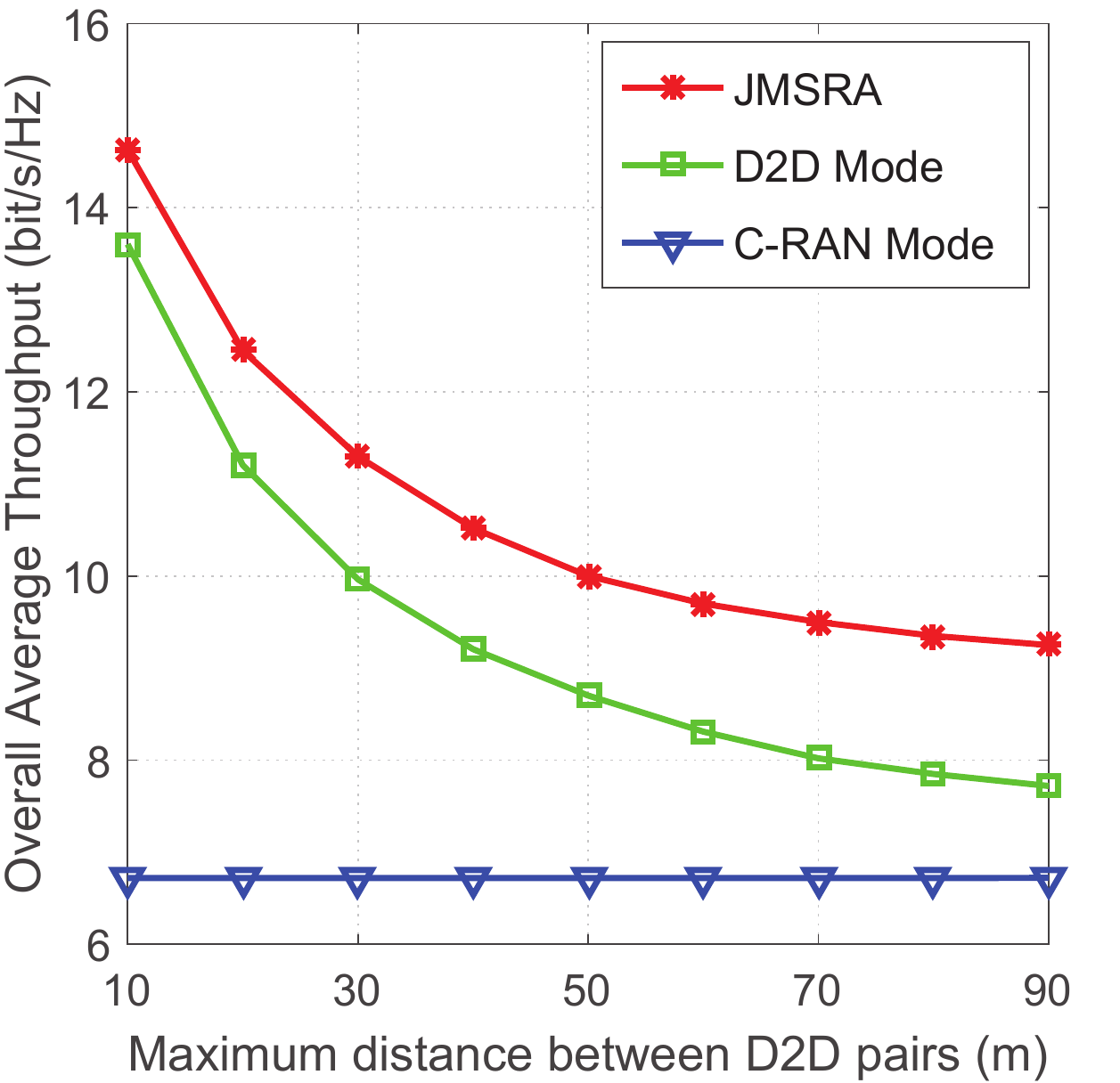}
\caption{Overall average throughput versus maximum distance between D2D pairs.}
\label{exp5}\vspace*{-1em}
\end{figure}
Fig. \ref{exp5} plots the overall average throughput of three algorithms against maximum distance between D2D pairs. From the figure, the overall average throughput for all algorithms except C-RAN Mode algorithm declines as the maximum distance between D2D pairs increases, which demonstrates the advantage of short-range D2D communications. For the proximity gain offered by D2D communications, it will decreases exponentially as the D2D link distance increases. In addition, the performance gap between the JMSRA algorithm and the D2D Mode algorithm becomes larger as the maximum distance between D2D pairs increases. This is because the advantage of C-RAN mode over D2D mode mainly comes from the large-scale collaborative signal processing, uplink CoMP implementation, and cooperative radio resource allocation provided by the BBU pool when the proximity gain offered by D2D communications is negligible.

\section{Conclusion}
As an evolution of heterogeneous ultra-dense network, the cloud radio access network (C-RAN) is severely constrained by the capacity-limited fronthaul and long end-to-end delay, and the device-to-device (D2D) is a good alternative technique. In this paper, we have focused on the stochastic optimization problem of resource allocation in C-RANs with D2D, by considering dynamic traffic arrivals and time-varying channel conditions. Lyapunov optimization technique has been utilized to
transform the the stochastic optimization problem into a delay-aware overall average throughput maximization problem, which is a mixed-integer nonlinear programming problem. To make this problem tractable, the optimization problem has been decomposed into three subproblems: mode selection, uplink beamforming design, and power control. The corresponding joint mode selection and resource allocation (JMSRA) algorithm based on the modified branch and bound method and weighted minimum mean square error approach has been proposed to solve these three subproblems iteratively. Simulation results have shown that the proposed JMSRA algorithm can quickly converge to a stationary point and have validated the good performance gain of the proposed JMSRA algorithm, which implies that C-RANs with D2D do have the advantages of achieving high throughput, reducing latency, and alleviating the burden on the constrained fronthaul. In addition, the proposed algorithm can approach a flexible average throughput-delay tradeoff on demand.


\appendices
\section{PROOF OF LEMMA 1}
We provide the proof for ease of understanding, as the results in this lemma are important for the algorithm design later.

Based on the fact that ${\left(\max[Q-R, 0]+A\right)}^2\leq Q^2+R^2+A^2-2Q(R-A), \forall Q,R,A\geq0$, squaring both sides of (\ref{queue}) yields
\begin{align}
{Q_k(t+1)}^2 \leq& {Q_k(t)}^2+{R_k(t)}^2+{A_k(t)}^2\nonumber\\
&-2Q_k(t)(R_k(t)-A_k(t)).
\end{align}

Summing over all $k (\forall k \in \mathcal{K})$ at both sides of the above equality and rearranging terms, we have
\begin{align}
&L(\bold{\Theta}(t+1))-L(\bold{\Theta}(t)) \nonumber\\
&\leq
\sum_{k\in \mathcal{K}}\frac{{R_k(t)}^2+{A_k(t)}^2}{2}-\sum_{k\in \mathcal{K}}Q_k(t)(R_k(t)-A_k(t)).
\end{align}

Taking conditional expectations and adding $-V\mathbb{E}\left\{\sum_{k\in \mathcal{K}}{R_k}(t)|\bold{\Theta}(t)\right\}$ at both sides of the above equality, there is
\begin{align}
\vartriangle & (\bold{\Theta}(t)) -V\mathbb{E}\left\{\sum_{k\in \mathcal{K}}{R_k}(t)|\bold{\Theta}(t)\right\}\nonumber\\
& \leq B+\sum_{k\in\mathcal{K}}{Q_k}(t)\mathbb{E}\{{A_k}(t)-{R_k}(t)|\bold{\Theta}(t)\}\nonumber\\
& \quad -V\mathbb{E}\left\{\sum_{k\in \mathcal{K}}{R_k(t)}|\bold{\Theta}(t)\right\},
\end{align}
where $B$ is a positive constant that satisfies
\begin{equation}
B\geq \frac{1}{2}\sum_{k\in \mathcal{K}}\mathbb{E}\left\{{R_k}(t)^2+{A_k}(t)^2|\bold{\Theta}(t)\right\}.
\end{equation}

\textit{Lemma 1} is proven.

\end{document}